\documentclass{ws-ijmpd}
\RequirePackage{graphicx}
\usepackage{epstopdf}
\usepackage{bm}
\usepackage{amsfonts}

\begin{document}

\title{Inflation in anisotropic brane universe using tachyon field}

\author{Rikpratik   Sengupta}
\address{Department of Physics, Government College of Engineering and Ceramic Technology, Kolkata 700 010, West Bengal, India \\ rikpratik.sengupta@gmail.com}

\author{Prasenjit Paul}
\address{Department of Physics, Government College of Engineering and Ceramic Technology, Kolkata 700 010, West Bengal, India \\Department of Physics, Indian Institute of Engineering Science and Technology, Howrah \\711 103, West Bengal, India \\prasenjit071083@gmail.com}

\author{Bikash Chandra Paul}
\address{Department of Physics, University of North Bengal, Siliguri 734 013, West Bengal, India \\ bcpaul@associates.iucaa.in}

\author{Saibal Ray}
\address{Department of Physics, Government College of Engineering and Ceramic Technology, Kolkata 700 010, West Bengal, India \\ saibal@associates.iucaa.in}
\maketitle

\begin{abstract}
Cosmological solution to the gravitational field equations in the generalized Randall-Sundrum model for an anisotropic brane with Bianchi I geometry and perfect fluid as matter sources has been considered. The matter on the brane is described by a tachyonic field. The solution admits inflationary era and at a later epoch the anisotropy of the universe washes out. We obtain two classes of cosmological scenario, in the first case universe evolves from singularity and in the second case universe expands without singularity.
\end{abstract}


\section{INTRODUCTION}
The idea of cosmological inflation~\cite{guth} is now largely accepted as a part of the standard model of big bang cosmology. It has been found capable of not only solving the flatness, horizon and relic problems \cite{sato1,sato2} but also supplying a mechanism for generating density perturbations~\cite{hawking,starobinsky,pi,steinhardt} which play a critical role in formation of large scale structures in the universe. On the observational front, inflation gets a good deal of support from the Cosmic Microwave Background Radiation (CMBR) and WMAP~\cite{bennett,peiris}  data. Within the framework of General Relativity (GR), the large scale uniformity of the universe in the presence of normal matter with positive pressure cannot be explained. The inflationary paradigm can successfully produces a FRW-like universe.

GR sometimes becomes inappropriate while dealing with situations where the energies are extremely high, e.g. in the case of the early universe. In such a situation it is generally believed that the theory of gravity has to be modified such that in the low energy limit we can get back the results obtained from standard GR. One such theory of modified gravity that has been used extensively to study the  high energy cosmological situations in the early universe is the brane world gravity~\cite{Maartens1}. One of the first brane world models was proposed by Randall and Sundrum (RS) in an attempt to solve the Hierarchy Problem~\cite{randall} in particle physics. Inspired by Superstring and M-theory, they considered that our universe is a (3+1) dimensional hypersurface (known as brane) embedded in a higher (5) dimensional space-time called “bulk” described by Anti-de Sitter (AdS) space characterized by a negative cosmological constant. The RS single brane model~\cite{sundrum} with a positive brane tension where the second negative tension brane is sent off to infinity, has been used extensively as a framework to study early universe cosmology~\cite{Maartens2,binetruy,Papa,Langlois}. It is considered that only the gravitational force is free to move into the bulk while the other natural forces are confined to the brane only. Interesting to note in the present context that inflation on the isotropic brane has been studied in details by Dvali and Tye~\cite{Dvali} and Sasaki~\cite{Sasaki}.

It is often sensible to consider anisotropic initial conditions for inflation~\cite{collins}. This issue has been addressed widely both in GR and also in the brane. The energy  distribution of an anisotropic Bianchi I universe has been studied by Radinschi~\cite{Radinschi}. Maartens et. al.~\cite{Maartens3} have explored the behaviour of an anisotropic Bianchi type-I brane world in the presence of a scalar field obtaining inflation even when large anisotropy is present, infact, the amount of inflation increases in the presence of anisotropy. A similar study has been done by Paul~\cite{paul1} where a new exact solution to the Einstein Field Equations (EFE) in RS brane with Bianchi type-I spacetime characterized by vanishing bulk Weyl tensor has been obtained. Due to the presence of a term quadratic in energy density in the modified EFE in this context, there is a modified behaviour of the early universe in contrast to the standard GR.

It has been shown by Sen~\cite{sen1} in the context of string theory that a pressureless dust like matter is produced by an unstable decaying D-brane. Such tachyon condensate matter has an interesting equation of state (EOS) which may be described by an effective field called {\it tachyon field} with a Lagrangian different from a scalar field. Such a field may be used to study the role of an {\it inflaton} - the field responsible for early inflation~\cite{sen2}. This type of attempts have been made by some authors~\cite{sami,paul2} for isotropic initial conditions.

In essence, in the present paper we consider anisotropic Bianchi type-I universe on the RS brane with a tachyon field and study the cosmological evolution of the brane at both low and high energies for either model considering a singular and a non-singular universe.

\section{EINSTEIN FIELD EQUATION ON THE ANISOTROPIC BRANE}
This section is basically a review of the gravitational field equations on the 3-brane~\cite{shiromizu} and it's extension to the consideration of the cosmological isotropic FRW brane~\cite{Maartens2,binetruy}  and more generally the anisotropic brane with Bianchi-I geometry~\cite{Maartens3,paul1}. The RS brane world model is inspired from the 11-dimensional theory of Horava and Witten~\cite{HW1996} on the orbifold $R^{10} X S^1/Z_2$ where the $Z_2$ or reflection symmetry is employed. In this Horava-Witten model standard model particles are confined to 4-dimensional spacetime but the gravitons, the quanta for the gravitational field have access to the full spacetime. RS simplied this to a 5-dimensional problem.  The study of the generalized RS brane in a cosmological context with the Friedmann metric was first performed by Binetruy et al.~\cite{binetruy} leading to a modification of the general relativistic Friedmann equation which contains only the first order term in the energy density. A fully geometric treatment of the RS brane model leading to the general modified EFE on the brane was first given by Shiromizu et al.~\cite{shiromizu}. 

The EFE in the 5-dimensional bulk is given as
\begin{equation}
G_{AB}^{(5)} = \kappa_{(5)}^{2} \left[ - g_{AB}^{(5)} \Lambda_{(5)} + 
T_{AB}^{(5)} \right], 
\end{equation}
where the energy-momentum tensor in the bulk is  $T_{AB}^{(5)} = \delta (y) [ - \lambda g_{AB} + T_{AB} ]$.

Using the junction condition of Israel~\cite{Israel1966} and considering $Z_2$ symmetry, implying the equality of quantities on the brane approaching from the + and - side, the modified EFE on the brane becomes
\begin{equation}
G_{\mu \nu} = - \Lambda g_{\mu \nu} + \kappa^{2} T_{\mu \nu}
+ \kappa_{(5)}^{4} S_{\mu \nu} - E_{\mu \nu}.
\end{equation}

The gravitational constant $k$ and effective cosmological constant $\Lambda$ on the four dimensional brane are respectively given by
\[
k^{2}  = \frac{1}{6} \lambda \kappa_{(5)}^{4},
\]
\begin{equation}
\Lambda = \frac{|\Lambda_{5}|}{2}  \left[  \left( \frac{\lambda}{\lambda_{c}} \right)^{2} - 1 \right], 
\end{equation}
where $\lambda_{c}$ is the critical brane tension which is defined as 
\begin{equation}
\lambda_{c} = 6 \frac{|\Lambda_{5}|}{\kappa_{5}^{2}} .
\end{equation}

Both the effective cosmological constant and gravitational constant on the brane are determined by the brane tension. In the original RS picture the $5$-dimensional bulk cosmological constant is fine tuned with the brane tension to give zero effective cosmological constant on the brane. However, in the generalized RS picture, as considered in our case, there is a non-zero cosmological constant on the brane.   

The major modifications to the EFE are encoded mainly in two terms:

$\bullet $
$S_{\mu \nu} $ leads to quadratic terms in the matter energy-momentum tensor given by
\begin{equation}
S_{\mu \nu} =  \frac{1}{12} T T_{\mu \nu} - \frac{1}{4}  T_{\mu \alpha} T^{\alpha}_{\nu} + \frac{1}{24} g_{\mu \nu} \left[ 3 T^{\alpha \beta} T_{\alpha \beta} - (T^{\alpha}_{\alpha})^{2} \right],
\end{equation}
where $T = T^{\alpha}_{\alpha}$. Here it is to be noted that $S_{\mu \nu}$ contributes only at high energies when $\rho > \lambda$. 

$\bullet$
$E_{\mu \nu} $ is the projection of the bulk Weyl tensor on the brane responsible for the transfer of gravitational effects from the bulk to the brane. We define a unit vector $n^A$ normal to the surface. The induced metric on the brane is given as $g_{AB}={g_{AB}}^{\left(5\right)}-n_{A}n_{B}$.
The Weyl projection on the brane is given as 
${E}_{\mu \nu} =  C^{(5)}_{ACBD} n^{C} n^{D} g_{\mu}^{A} g_{\nu}^{B}$ which is symmetric and traceless. It has no components orthogonal to the brane. 

All the local and the non-local bulk corrections may be written in a compact form where the modified EFEs take the standard form:
\begin{equation}
G_{\mu \nu} = - \Lambda g_{\mu \nu} + \kappa^{2} T_{\mu \nu}^{total},
\end{equation}
where $ T_{\mu \nu}^{total} = T_{\mu \nu} + \frac{6}{\lambda} S_{\mu \nu} - \frac{1}{\kappa^{2}} E_{\mu \nu}$.

We take $u^{\mu}$ to be the four-velocity comoving with matter. The projection into the comoving rest space is given as $h_{\mu \nu}=g_{\mu \nu}+u_{\mu} u_{\nu}$.

The effective total energy density, pressure, anisotropic stress and energy flux are~\cite{Maartens2} 
\[
\rho^{total} = \rho \left(  1 + \frac{\rho}{2 \lambda} \right) + \frac {6{U}}{\kappa^{4} \lambda },
\]
\[
p^{total} = p + \frac{\rho}{2 \lambda} (\rho + 2 p) + \frac {2{U}}{\kappa^{4} \lambda },
\]
\[
\pi^{total} = \frac {6 }{\kappa^{4} \lambda } {P_{\mu \nu}}, 
\]
\begin{equation}
q^{total}_{\mu} = \frac {6 }{\kappa^{4} \lambda } {Q_{\mu \nu}},
\end{equation}
where 
${U}=-\frac{\kappa^2 \lambda E_{\mu \nu} u^{\mu} u^{\nu} }{6}$ is the effective non-local energy density on the brane,
${P_{\mu \nu}}=-\frac{\kappa^2 \lambda \left(h_{\mu}^{\alpha}h_{\nu}^{\beta}-\frac{1}{3}h^{\alpha \beta}h_{\mu\nu}\right)E_{\alpha \beta}}{6}$ is the effective non-local anisotropic stress and $ {Q_{\mu}=\frac{\kappa^2 \lambda h_{\mu}^{\alpha} E_{\alpha \beta} u^{\beta} }{6}}$ is the effective non-local energy flux on the brane that carries Coulomb and gravito-magnetic effects from the free gravitational field in the bulk~\cite{Maartens2,Maartens3}.

 The brane energy-momentum tensor and the overall effective energy momentum tensor are both conserved separately. Here we shall consider the case of a Bianchi-I brane geometry which is a simple generalization of Friedmann-Robertson-Walker (FRW) brane geometry introducing anisotropy in the geometry.

An anisotropic brane-world model with Bianchi type-I space is considered. The Bianchi-I brane has the induced metric:
\begin{equation}
ds^{2} =  -  dt^{2} +  \sum_{i=1}^{3} R_{i}^{2} (dx^{i})^{2},
\label{metric}\end{equation}
and is characterized by
\[
D_{\mu} f = 0, \; A_{\mu} = \omega_{\mu} = 0, \; {\bf Q_{\mu}} = 0 ,\; 
R_{\mu \nu}^{*} = 0,
\]
where $ D_{\mu} $ is the projected covariant spatial derivative, $f$ is any physically defined scalar, $A_{\mu } $ is the four-acceleration, $\omega_{\mu} $  is the vorticity and $R_{\mu \nu}^{*} = 0$ is the Ricci tensor of the 3-surface orthogonal to $u^{\mu}$.

The conservation equations reduce to 
\[
\dot{\rho} + \Theta (\rho + p) = 0,
\]
\[
\dot{{\bf U}} +\frac{4}{3} \Theta {\bf U} + \sigma^{\mu \nu} P_{\mu \nu} = 0,
\]
\begin{equation}
D^{\nu} P_{\mu \nu} = 0,
\end{equation}
where a dot denotes $u^{\nu} \bigtriangledown_{\nu}$, $\Theta$ is the volume expansion rate, $\sigma_{\mu \nu} $ is the shear. 

The directional Hubble parameters are defined by
\[
H_{i} = \frac{\dot{R_{i}}}{R_{i}},
\]
and the mean expansion factor $a = \left( R_{1} R_{2} R_{3} \right)^{\frac{1}{3}} $. One gets the expansion rate 
\begin{equation}
\Theta = 3 H = 3 \frac{\dot{a}}{a} = \sum_{i=1}^{3} H_{i}.
\end{equation}

The average anisotropy expansion is given by
\begin{equation}
A = \frac{1}{3} \sum_{i = 1}^{3} \left (\frac{\triangle H_{i}}{H}  \right)^{2}.
\end{equation}

We are not able to obtain the evolution equation for $P_{\mu \nu}$  as there can be no closed system of equations on the brane, i.e. without the bulk equations, the brane dynamics cannot be determined. For the bulk geometry to be conformally flat as in the case of FRW branes, we set ${U}  = 0$ as $\sigma^{\mu \nu} P_{\mu \nu} = 0$.

For the metric (\ref{metric}), using vanishing of $R_{\mu \nu}^{*}$, it is obtained via integrating the shear-contracted Gauss-Codazzi equation on the brane that:  
\begin{equation}
\sigma^{\mu \nu} \sigma_{\mu \nu} = \sum_{i=1}^{3} (H_{i} - H)^{2} = \frac{6 \Sigma^{2}}{a^{6}} , \; \; \;  \dot{\Sigma } = 0,
\label{shear metric} 
\end{equation}
and the  anisotropy parameter becomes
\begin{equation}
A = \frac{\Sigma^{2}}{H^{2} V^{2}},
\label{anisotropy}
\end{equation}
where $V = a^{3}$.

We now use equation (\ref{shear metric}) to obtain generalized Friedmann equation for the Bianchi-I brane which is  given by 
\begin{equation}
H^{2} = \frac{8 \pi G}{3}  \rho \left( 
1 + \frac{\rho}{2 \lambda} \right)  + \frac{\Sigma^{2}}{a^{6}} + \frac{\Lambda}{3}.
\label{friedmann}
\end{equation}

We replace here $\kappa^{2} $ by $8 \pi G$. One recovers the field equation corresponding to the FRW brane when anisotropy $\Sigma = 0$ \cite{binetruy}, in the presence of a cosmological constant $\Lambda$.

\section{COSMOLOGICAL SOLUTION}
The matter content in the universe is a minimally coupled tachyon field $\psi = \psi(t)$ on the brane.

The energy momentum tensor describing the tachyon field on the brane is given by
\begin{equation}
\label{3}
T_{\mu\nu}=-\frac{V{(\psi)}}{\sqrt{1-\partial_{\alpha}\psi\partial^{\alpha}\psi}}\,\partial_{\mu}\psi\partial_{\nu}\psi
+g_{\mu\nu}\Big(-V(\psi)\sqrt{1-\partial_{\alpha}\psi\partial^{\alpha}\psi}\Big),
\end{equation} 
where the energy density and pressure~\cite{paul3} are obtained as
\begin{equation}
\label{4} \rho_{\psi}=\frac{V(\psi)}{\sqrt{1-\dot{\psi}^{2}}},
\end{equation}
and
\begin{equation}
\label{5} p_{\psi}=-V(\psi)\sqrt{1-\dot{\psi}^{2}},
\end{equation}
where $\frac{1}{2}{\dot{\psi}}^2$ and $V(\psi)$ represent the kinetic and potential energies respectively.

The equation of motion for the tachyon field~\cite{paul3} is given as 
 \begin{equation}
 \label{8}
 \frac{\ddot{\psi}}{1-\dot{\psi}^{2}}+3H\dot{\psi}+\frac{V'}{V}=0,
 \end{equation} 
where `over dot' and 'prime' are denoting differentiation with respect to time and $\psi$ respectively.
 
 In order to obtain a cosmological solution we consider the simplest case of a tachyon field with constant potential $V(\psi)=V_0=$constant. Putting this in Eq. (\ref{8}), we obtain 
 \begin{equation}
 \dot{\psi}=\frac{1}{\sqrt{1+Ca^6}},
 \label{9}
 \end{equation} 
 where $C$ is the constant of integration. The first Friedmann equation on the brane can now be obtained by plugging the value of $\rho$ in Eq. (\ref{friedmann}) in terms of the scale factor using Eqs. (\ref{4}) and (\ref{9}) as 
 \begin{equation} 
 H^2=\left(\frac{\dot a}{a}\right)^2= H_0^2+ \frac{\eta}{a^3}+\frac{\eta_1}{a^6},
 \end{equation}
 where $ H_0^2= \frac{\Lambda}{3}+\frac{\kappa^2V_0^2}{6\lambda},~\eta=\pm\frac{\kappa^2V_0}{3\sqrt{C}}$ and $\eta_1=\frac{\kappa^2V_0^2}{6\lambda C}+\Sigma^2$. We shall consider solutions with only the positive root at first, calculating the corresponding anisotropy parameter $A$ and tachyon field $\psi$ for the obtained solution. Then we shall study the modification in the anisotropy on replacing the positive root by the negative root in the final expression for $A$.
 
 We have assumed that $\sqrt{1+Ca^6}$$\approx$1. This approximation is reasonable as in the very early universe, before inflation, the scale factor $a(t)$ is very small. In order to obtain the solution for the scale factor we substitute $y=\ln a(t)$. This gives from Eq. (20) as follows:
 \begin{equation}
 \frac{dy}{dt} = \pm \sqrt{ H_0^2 + \eta e^{-3y}+  \eta_{1}  e^{-6y}}.
 \end{equation}
 
 Solving the above differential equation the scale factor is obtained as 
 \begin{equation}
 a(t)= \left[\frac{B}{4H_0^2}e^{3H_0 t}+\frac{\eta^2-4\eta_1 H_0^2}{4BH_0^2}e^{-3H_0 t}-\frac{\eta}{2H_0^2}\right]^{\frac{1}{3}},
 \label{gen scale}
 \end{equation}
 where $B$ is the integration constant and the scale factor depends on the brane tension $\lambda$. Depending on the values of the integration constant $B$, which can be determined from the boundary condition we consider two different cases of singular and non-singular universe.

 \subsection{Singular universe} 
 Singular solution is obtained for $B_{\pm}$=$\eta \pm 2H_0\sqrt{\eta_1}$. The corresponding scale factor for integration constant $B_+$ has the form 
  \begin{equation}
 a(t)= \left[\frac{\eta + 2H_0\sqrt{\eta_1}}{4H_0^2}e^{3H_0 t}+\frac{\eta^2-4\eta_1 H_0^2}{4(\eta + 2H_0\sqrt{\eta_1})H_0^2}e^{-3H_0 t}-\frac{\eta}{2H_0^2}\right]^{\frac{1}{3}}.
 \end{equation}
 
 We now study the solution for both low energy limit (GR) as well as high energy limit (brane) in the next subsection.
 
 \subsubsection{Low energy limit}
 GR from the RS brane model is revisited when we consider $\frac{V_0}{\lambda}\to 0$. In this limit the coefficients of the terms in the scale factor take the form as follows: $ H_0^2= \frac{\Lambda}{3},~\eta=\frac{\kappa^2V_0}{3\sqrt{C}}$ and $\eta_1=\Sigma^2$.
 
 The scale factor for the corresponding coefficients in the low energy limit turns out to be 
 \begin{equation}
 a(t)=\left[\left(\frac{\kappa^2 V_0}{12\sqrt{C}H_0^2}+\frac{\Sigma}{2H_0}\right)e^{3H_0 t}+\left(\frac{\kappa^2 V_0}{12\sqrt{C}H_0^2}-\frac{\Sigma}{2H_0}\right)e^{-3H_0 t}-\frac{\kappa^2 V_0}{6\sqrt{C}H_0^2}\right]^{\frac{1}{3}}.
 \end{equation}
 
 Putting the above value of scale factor in Eq. (\ref{9}) we get the time variation of the tachyon field $\psi$ as  
 \begin{equation}
 \dot{\psi}= \frac{1}{\sqrt{1+\left[\left(\frac{\kappa^2 V_0}{12H_0^2}+\frac{\Sigma\sqrt{C}}{2H_0}\right)e^{3H_0 t}+\left(\frac{\kappa^2 V_0}{12H_0^2}-\frac{\Sigma\sqrt{C}}{2H_0}\right)e^{-3H_0 t}-\frac{\kappa^2 V_0}{6H_0^2}\right]^2}}.
 \end{equation}
 
The above equation is not integrable in general. So, in order to get a solution for the tachyon field we consider the case in which the term inside the bracket in the denominator of the equation dominates. The solution of $\psi$ can be written as 
 \begin{equation}
 \psi=\psi_0 -\frac{2}{3\sqrt{C}\Sigma}\tanh^{-1}\left[\frac{\left(\frac{\kappa^2V_0}{6H_0^2}+\frac{\Sigma\sqrt{C}}{H_0}\right)e^{3H_0 t}-\frac{\kappa^2V_0}{6H_0^2}}{\frac{\Sigma\sqrt{C}}{H_0}}\right].
 \label{psi-aniso}
 \end{equation}
  
 The anisotropy parameter as given in (\ref{anisotropy}) turns out to be  
 \begin{equation}
 A = \frac{9\Sigma^{2}}{[\frac{\kappa^{2}V_0}{2\sqrt{C}H_0}\sinh(3 H_0 t) + 3\Sigma \cosh(3 H_0 t)]^2}.
 \end{equation} 
 
 The variation of the anisotropy parameter $A(t)$ with time for this case is shown in Fig. 1.

 Now taking zero anisotropy we set $\Sigma=0$. In this case the scale factor becomes 
 \begin{equation}
 a(t)=\left[\frac{\kappa^{2}V_0}{6\sqrt{C}H_0^2}\right]^{\frac{1}{3}} \sinh^{\frac{2}{3}}(\frac{3}{2} H_0 t).
 \end{equation}

The corresponding time variation of the tachyon field is obtained as
\begin{equation}
\dot{\psi}=\frac{1}{\left[\sqrt{1+\left\{\frac{\kappa^2V_0C^\frac{5}{2}}{6H_0^2}\right\}^\frac{1}{3} \sinh^{4} (\frac{3}{2} H_0 t)}\right]}.
\end{equation} 

 The above equation again in general is not integrable. So, in order to get a solution for the tachyon field we consider the case in which the term 1 in the denominator of the above equation is very small w.r.t. the other term under root. Then the solution can be written as
\begin{equation}
\psi=\psi_0 -\frac{\coth(\frac{3H_0t}{2})}{\left({\frac{\kappa^2V_0C^\frac{5}{2}}{6H_0^2}}\right)^\frac{1}{3}\frac{3H_0}{2}}.
\label{psi-iso}
\end{equation}

\begin{figure}[t]
	\centering
	\includegraphics[scale=0.4]{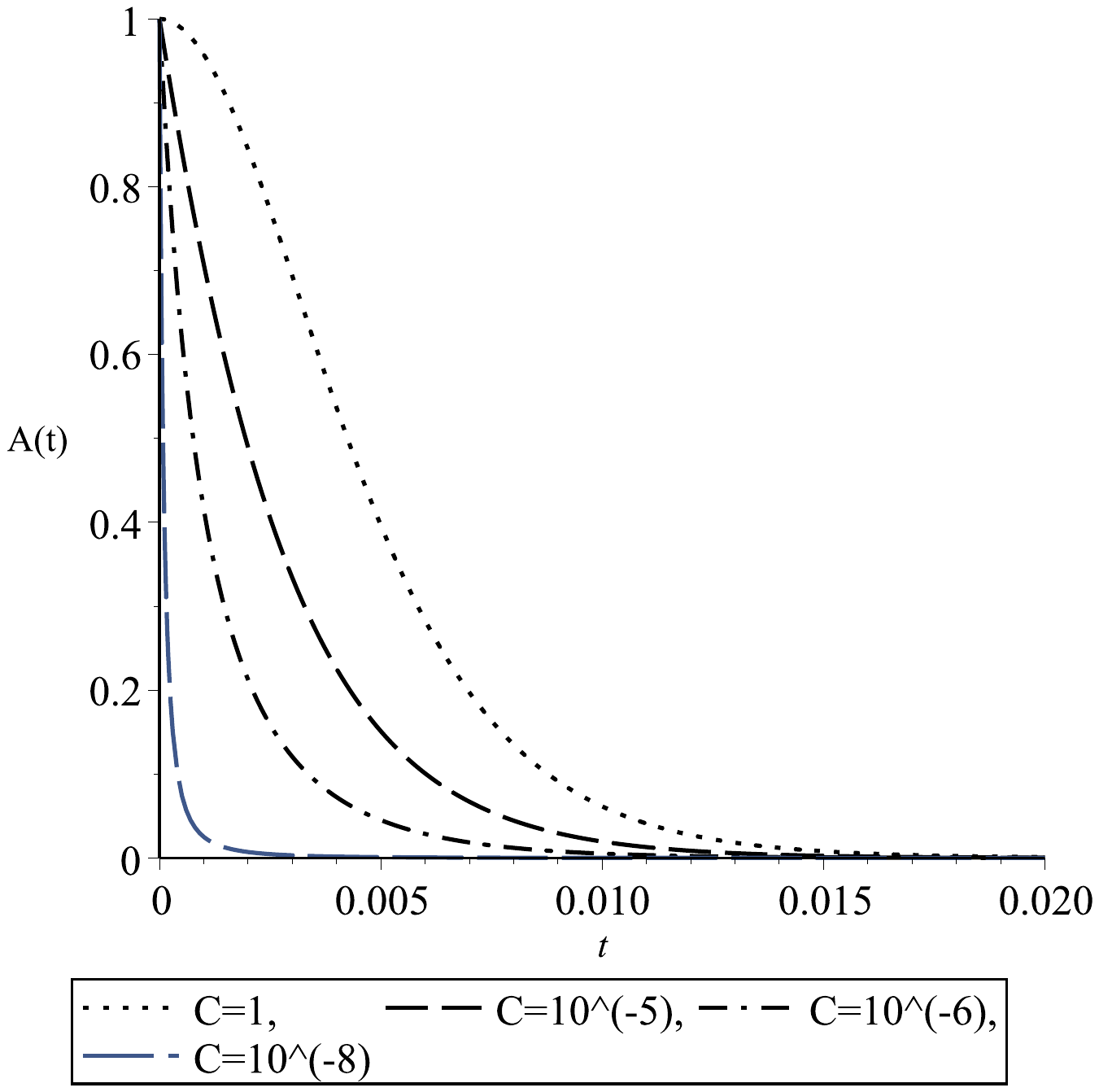}~\hspace*{-3.5cm}
	\includegraphics[scale=0.4]{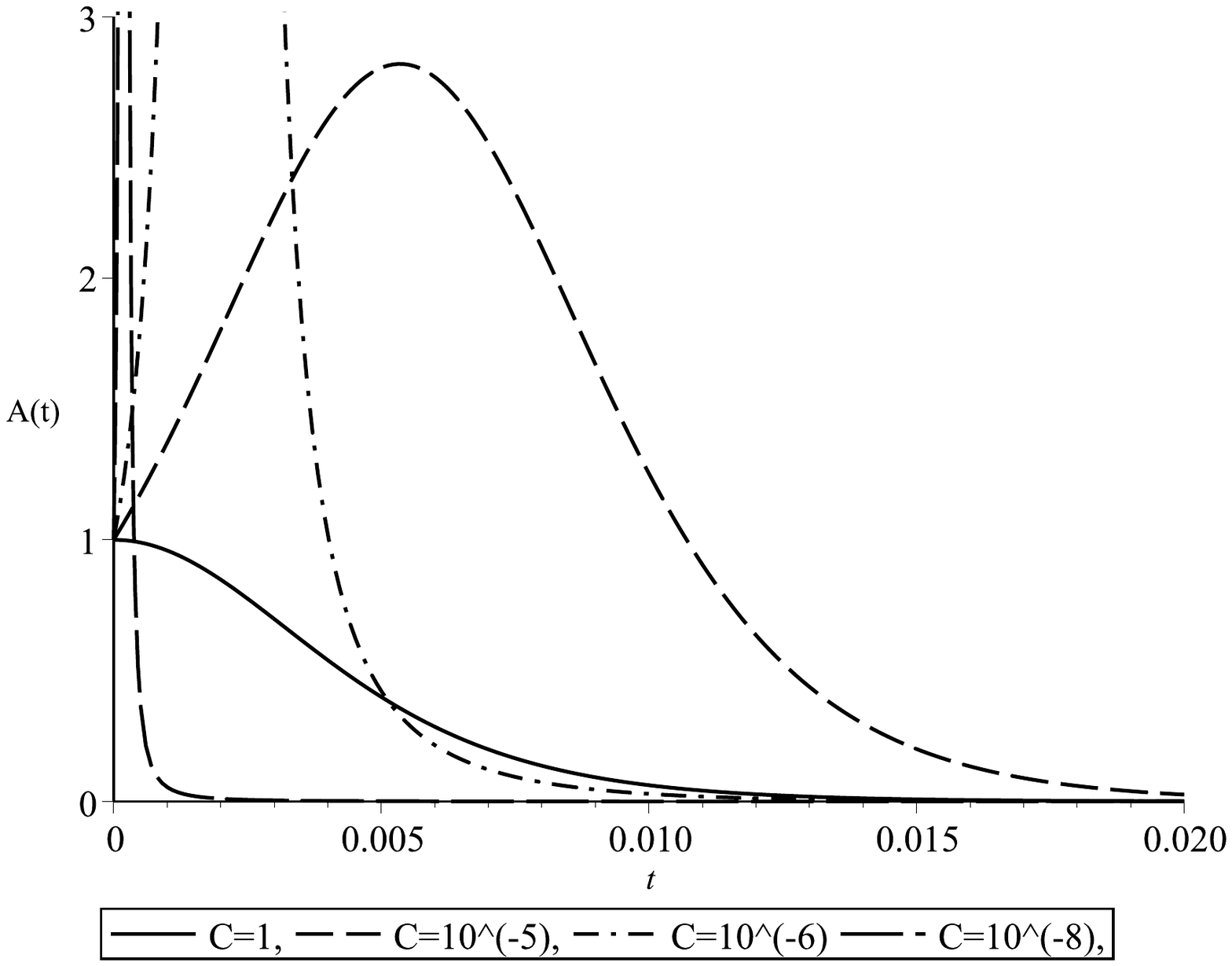}
     \vspace*{-5cm}
	\caption{$A(t)$ vs $t$ for Singular low energy case $ + \sqrt{C}$ (left panel) and $ - \sqrt{C}$ (right panel)} \label{fig:singlow+}
\end{figure}

\subsubsection{High energy limit} 
In this limit the brane effects become important, when $\frac{V_0}{\lambda}\to \infty$.

The coefficients of the scale factor turn out to have the form,
$ H_0^2= \frac{\kappa^2V_0^2}{6\lambda}, \eta=\frac{\kappa^2V_0}{3\sqrt{C}}$ and $\eta_1=\frac{\kappa^2V_0^2}{6\lambda C}+\Sigma^2$.

As $\frac{V_0}{\lambda}\to \infty$ so the negative exponential term becomes vanishingly small.

The scale factor a(t) is given here by
 $a(t)= \left[\frac{\eta + 2H_0\sqrt{\eta_1}}{4H_0^2}e^{3H_0 t}-\frac{\eta}{2H_0^2}\right]^{\frac{1}{3}}$, which finally can be expressed as 
 \begin{equation}
 a(t)=\left[\left(\frac{\lambda}{2\sqrt{C}V_0}+\sqrt{\frac{1}{4C}+\frac{3\lambda\Sigma^2}{2\kappa^2V_0^2}}\right)e^{3H_0 t}-\frac{\lambda}{\sqrt{C}V_0}\right]^\frac{1}{3}.
 \end{equation}

 The anisotropy parameter ($A$) is given in this case by the equation
\begin{equation}
A=\frac{\Sigma^2}{H_{0}^2\left({\frac{\lambda}{2\sqrt{C}V_{0}}+\sqrt{\frac{1}{4C}+\frac{3\lambda \Sigma^2}{2\kappa^2V_{0}^2}}}\right)^2\exp\left({6H_{0}t}\right)}.
\end{equation}

From the above equation it is evident that there will be inflation even with a large initial anisotropy. The variation of the anisotropy parameter $A(t)$ with time for this case is shown in Fig. 2.

We can consider two possible cases here:

Case (i) $\frac{1}{4C} \ll \frac{3\lambda\Sigma^2}{2\kappa^2V_{0}^2} $ which implies
         $ \Sigma^2 \gg \frac{\kappa^2V_{0}^2}{6\lambda C}.$

In this case using the initial condition for a singular universe $a(t=0)=0$, we obtain $\Sigma^2=\frac{\kappa^2 \lambda}{6C}$, which leads the constraint $ \frac{V_{0}^2}{\lambda^2} \ll 1.$ 

Case (ii) $\frac{1}{4C} \gg \frac{3\lambda\Sigma^2}{2\kappa^2V_{0}^2} $ which implies
          $ \Sigma^2 \ll \frac{\kappa^2V_{0}^2}{6\lambda C}.$
     
In this case using the initial condition for a singular universe $a(t=0)=0$, we obtain the condition $V_{0}^2=\lambda^2$.

If we consider $\Sigma^2 \ll \frac{\kappa^2C^2V_0}{3\lambda}$, the scale factor is given by the simple de Sitter solution
\begin{equation}
a(t)= \frac{1}{(4C)^\frac{1}{6}}e^{H_0 t}.
\end{equation}
 
  The corresponding tachyon field obtained by putting the above expression for the scale factor in Eq. (\ref{9}) is 
  \begin{equation}
  \psi=\psi_0 -\frac{\tanh^{-1}\left(\frac{1}{2}\sqrt{4+e^{6H_0 t}}\right)}{3H_0}.
  \end{equation}
  
\begin{figure}[t]
	\centering
	\includegraphics[scale=0.4]{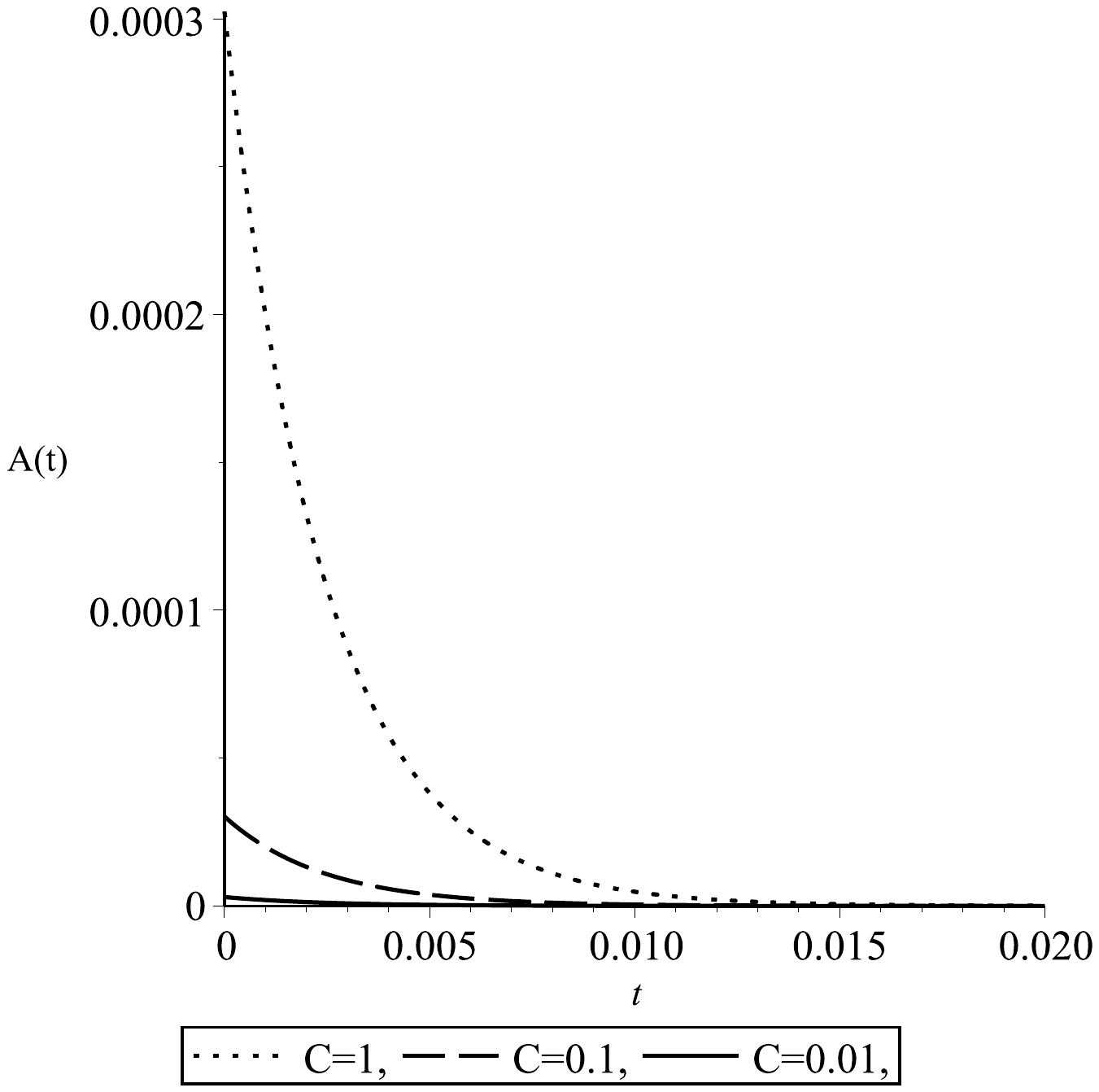}~\hspace*{-3.5cm}
	\includegraphics[scale=0.4]{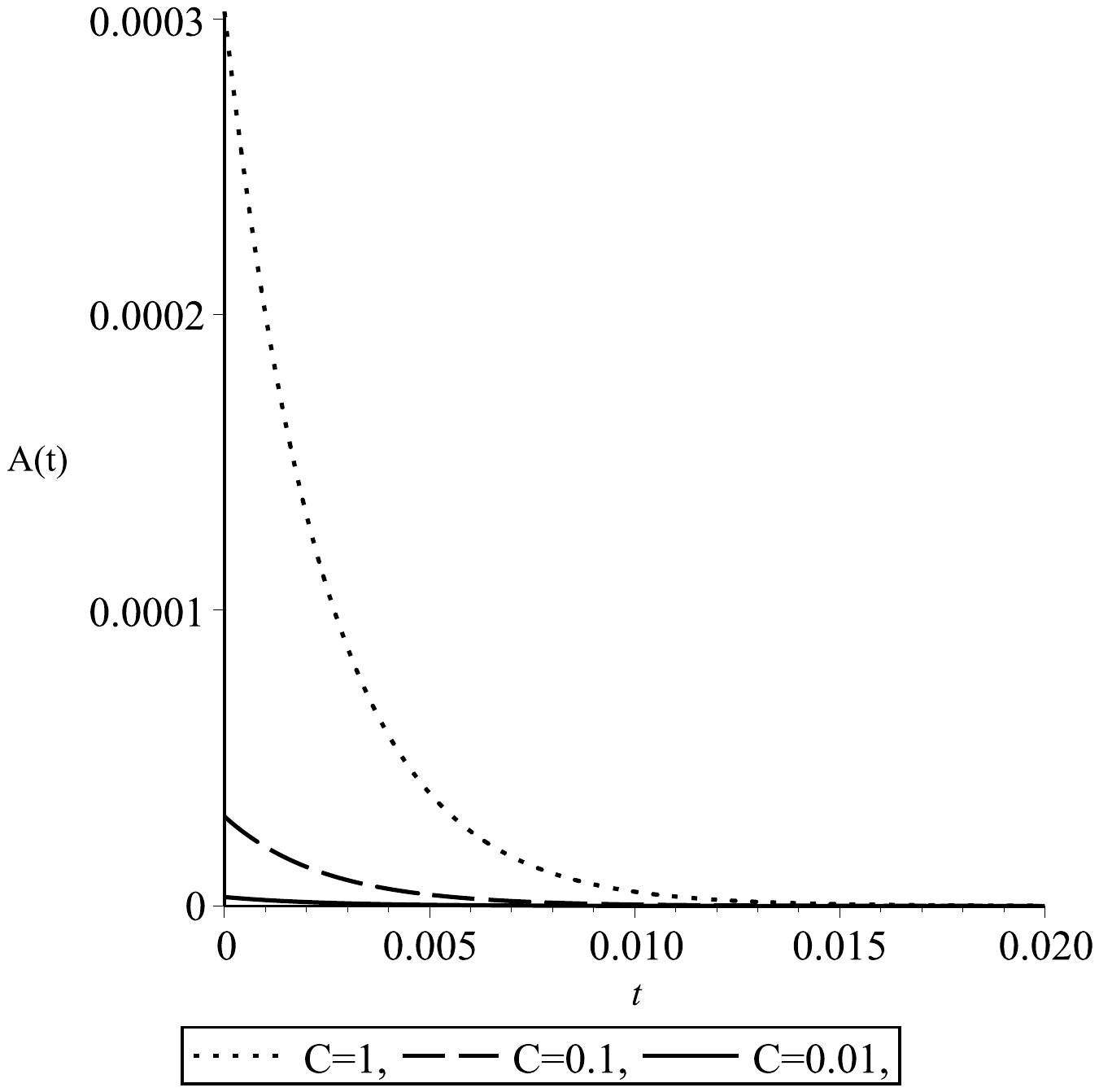}
	\vspace*{-5cm}
	\caption{$A(t)$ vs $t$ for Singular high energy case $ + \sqrt{C}$ (left panel) and $ - \sqrt{C}$ (right panel)} \label{fig:singhigh+}
\end{figure}
 
  Now we consider an alternative solution in the high energy limit for the integration constant $ B_-= \eta-2H_0\sqrt{\eta_{1}}=\frac{\kappa^2V_0}{3\sqrt{C}}\left[1-\frac{2V_0}{\lambda}-\frac{3\Sigma^2C}{\kappa^2V_0}\right].$
  
  This gives a corresponding scale factor  
  \begin{equation}
  a(t)= \left[\frac{\kappa\lambda}{\sqrt{C}}\left\{\frac{1}{2}\left(1-\frac{V_0}{\lambda}-\frac{3\Sigma^2C}{\kappa^2V_0}\right)e^{3H_0 t}-1\right\}\right]^\frac{1}{3}.
  \end{equation}
  
  The corresponding anisotropy parameter in this case is given by the equation  
   \begin{equation}
   A=\frac{4C\Sigma^2}{H_{0}^2\kappa^2 \lambda^2 \left(1-\frac{V_{0}}{\lambda}-\frac{3\Sigma^2C}{\kappa^2 V_{0}}\right)^2 \exp\left(6H_{0}t\right)}.  
   \end{equation}
   
   The variation of the anisotropy parameter $A(t)$ with time for this case is shown in Fig. 3.
   
   \begin{figure}[t]
   	\centering
   	\includegraphics[scale=0.4]{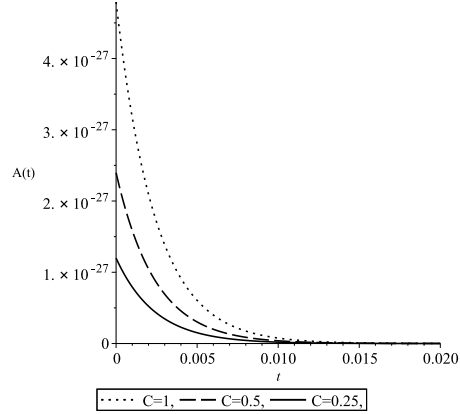}\\
   	\vspace{-5cm}
   	\caption{$A(t)$ vs $t$ for Singular high energy case $B-,\pm\sqrt{c}$}\label{fig:singhigh+-}
   \end{figure}
   
  In this case also two possible situations can be considered using the singular initial condition:
  
Case (i): $ \frac{V_{0}}{\lambda} \ll \frac{3\Sigma^2 C}{\kappa^2 V_{0}} $
   which implies 
    $ \Sigma^2 \gg \frac{\kappa^2V_{0}^2}{3\lambda C}.$
    
    In this case using the initial condition for a singular universe $a(t=0)=0$, we obtain $\Sigma^2=\frac{\kappa^2 V_{0}}{3C}$,which leads the constraint $ \frac{V_{0}}{\lambda} \ll 1,$ same as the one obtained in the case for $B_{+},$ taking the positive root only.
    
Case (ii): $ \frac{V_{0}}{\lambda} \gg \frac{3\Sigma^2 C}{\kappa^2 V_{0}} $  
    which implies
      $ \Sigma^2 \ll \frac{\kappa^2V_{0}^2}{3\lambda C}. $
 
    In this case using the initial condition for a singular universe $a(t=0)=0$, we obtain the condition $V_{0}=\lambda$, which is again same as the one obtained in the case for $B_{+},$ taking the positive root only.
    
    In the RS model the brane tension is taken positive for the single brane model and hence for $B_{+}$ the constant potential may be either positive or negative but for  $B_{-}$ the constant potential must always be positive.
    
\subsection{Non-singular universe}
For a non-singular universe with $a(t=0)=a_0$, the integration constant $B$ has the form
\begin{equation}
B=2H_0^2a_0^3+\eta \pm 2H_0\sqrt{H_0^2a_0^6+\eta_0^3+\eta_{1}}.
\end{equation}

 We obtain a simplified solution with the condition, $\eta^2=4\eta_{1}H_0^2$. Applying this condition, with the above value of integration constant $B$, the scale factor (\ref{gen scale}) takes the form 
 \begin{equation}
 a(t)=\left[\left(a_0^3+\frac{\eta}{2H_0^2}\right)e^{3H_0 t}-\frac{\eta}{2H_0^2}\right]^\frac{1}{3}.
 \end{equation}

\subsubsection{Low energy limit}
In this limit the values of the coefficients remain same as in the case of singular universe. The scale factor in the low energy limit takes the form
\begin{equation}
a(t)=\left[\left(a_0^3+\frac{\kappa^2V_0}{6\sqrt{C}H_0^2}\right)e^{3H_0 t}-\frac{\kappa^2V_0}{6\sqrt{C}H_0^2}\right]^\frac{1}{3}.
\end{equation}

The anisotropy parameter $A$ is given by 
\begin{equation}
A=\frac{\Sigma^2}{{H_{0}^2}\left(a_{0}^3+\frac{\kappa^2 V_{0}}{6 \sqrt{C}H_{0}^2}\right)^2\exp\left({6H_{0}t}\right)}.
\end{equation}

There is inflation despite large initial anisotropy and the anisotropy is washed out exponentially with time. The variation of the anisotropy parameter $A(t)$ with time for this case is shown in Fig. 4.  The corresponding tachyon field has the form 
\begin{eqnarray}
 \psi(t) = & & \psi_0-\frac{1}{\sqrt{\frac{3}{2}(\kappa^2V_0+6H_0^2)}}\ln\Bigg[\frac{1}{e^{3H_0 t}}\bigg\{\frac{\kappa^4V_0^2}{18H_0^4}+2+\left(-2\sqrt{C}a_0^3\frac{\kappa^2V_0}{6H_0^2}-\frac{\kappa^4V_0^2}{18H_0^4}\right)e^{3H_0 t}\nonumber \\
& & +2\Big[\big( {1+\frac{\kappa^4V_0^2}{36H_0^4}}\big)  \left(Ca_0^6+\frac{2a_0^3\sqrt{C}\kappa V_0}{6H_0^2}+\frac{\kappa^4V_0^2}{36H_0^4}\right)e^{6H_0 t}+\left(\frac{2a_0^3\sqrt{C}\kappa V_0}{6H_0^2}-\frac{\kappa^4V_0^2}{18H_0^4}\right)e^{3H_0 t}\nonumber \\
	& &+\frac{\kappa^4V_0^2}{36H_0^4}+1\Big]^{\frac{1}{2}}\bigg\}\Bigg].
\end{eqnarray}

It is to be noted that the cosmological solution is free of anisotropy but depends on the brane tension.

\begin{figure}[h]
	\centering
	\includegraphics[scale=0.4]{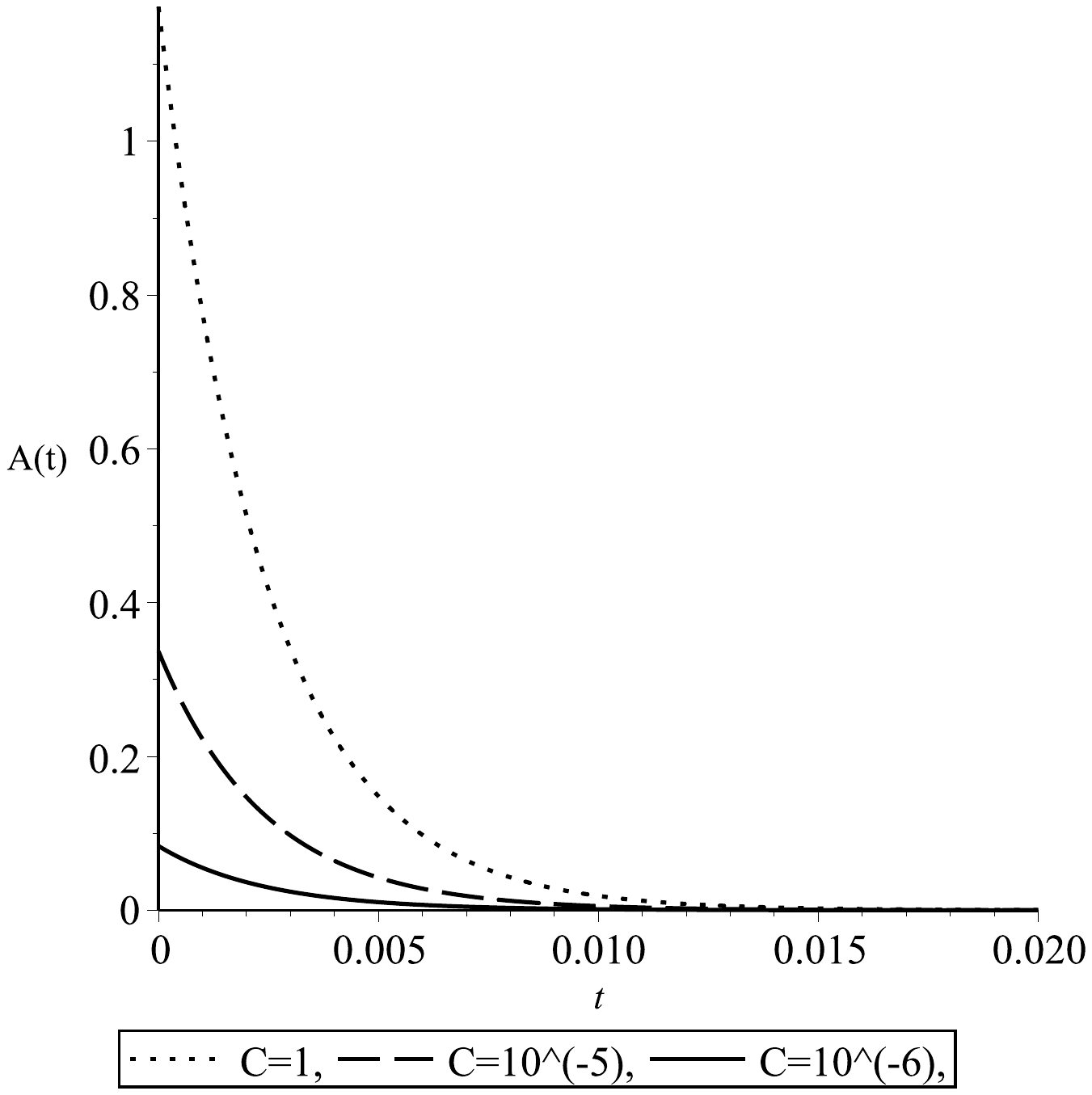}~\hspace*{-3.5cm}
	\includegraphics[scale=0.4]{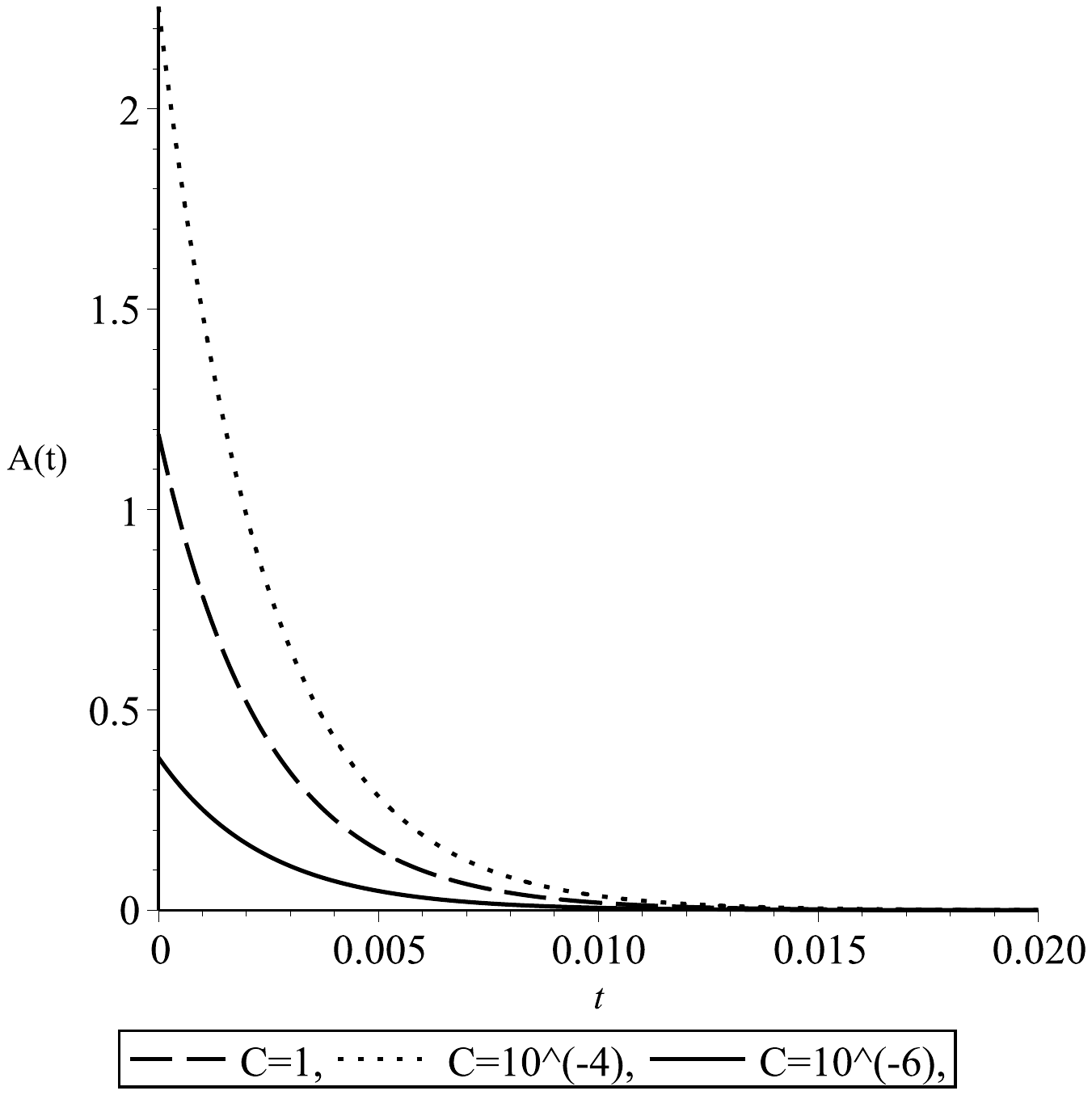}
	\vspace*{-5cm}
	\caption{$A(t)$ vs $t$ for Nonsingular low energy case $ + \sqrt{C}$ (left panel) and $ - \sqrt{C}$ (right panel)} \label{fig:nonsinglow}
\end{figure}

\subsubsection{High energy limit}
In this limit also the coefficients have the same form as in the case of singular universe in high energy limit. The scale factor, again found to be free of anisotropy as in the low energy limit turns out to be
\begin{equation}
a(t)=\left[\left(a_0^3+\frac{\lambda}{\sqrt{C}V_0}\right)e^{3H_0 t}-\frac{\lambda}{\sqrt{C}V_0}\right]^\frac{1}{3}.
\end{equation}

The anisotropy parameter $A$ is given by 
\begin{equation}
A=\frac{\Sigma^2}{{H_{0}^2}\left(a_{0}^3+\frac{\lambda}{ \sqrt{C}V_{0}}\right)^2\exp\left({6H_{0}t}\right)}.
\end{equation}

The variation of the anisotropy parameter $A(t)$ with time for this case is shown in Fig. 5.

\begin{figure}[h]
	\centering
	\includegraphics[scale=0.4]{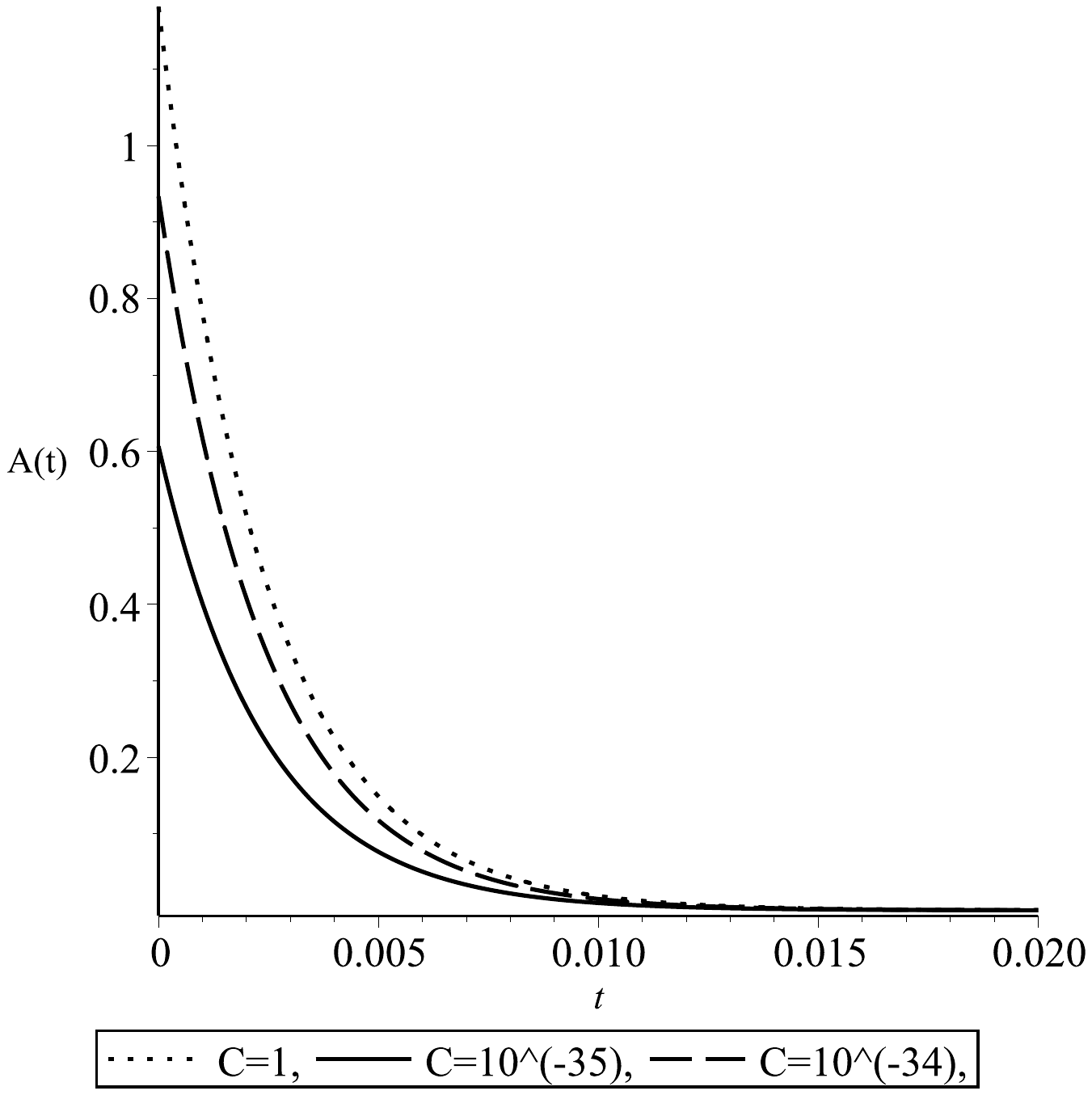}~\hspace*{-3.5cm}
	\includegraphics[scale=0.4]{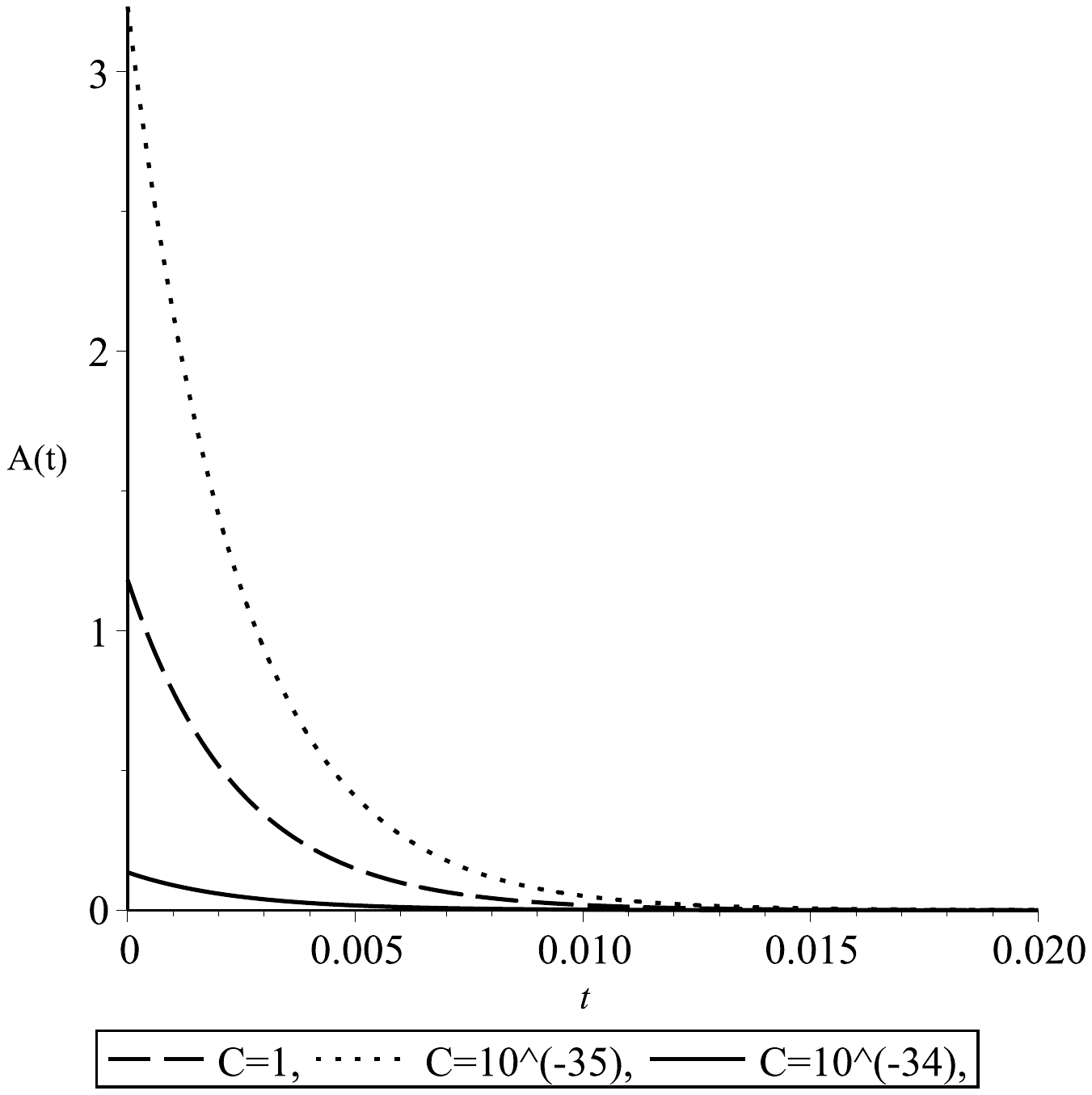}
	\vspace*{-5cm}
	\caption{$A(t)$ vs $t$ for Nonsingular high energy case $ + \sqrt{C}$ (left panel) and $ - \sqrt{C}$ (right panel)} \label{fig:nonsinghigh}
\end{figure}

There is inflation despite large initial anisotropy and the anisotropy is washed out exponentially with time just like in the low energy limit.

\section{DISCUSSIONS AND CONCLUSION}
We obtain a class of cosmological solutions for Bianchi-I geometry in generalized RS-II brane model with non-zero cosmological constant on the brane. We have studied inflation with a tachyon field, first in an universe evolving from a singularity and then in a non-singular universe. In both the cases the potential for the tachyonic field is taken as constant. The solutions are obtained in the low and high energy limiting conditions corresponding to the  dominance of GR and brane effects respectively. 

We have plotted the evolution of the anisotropy parameter with time and we find that in each case the anisotropy parameter drops down to almost zero value implying the washing out of the anisotropy resulting in an almost isotropic universe as observed today. As the value of the constant $C$ is lowered, the anisotropy parameter drops to zero more rapidly at earlier times. The evolution is however not identical for both positive and negative value of $\eta$. 

In the singular low energy case there is a clear difference in the dropping of the anisotropy parameter as compared to the other cases, as in this case the anisotropy parameter first rises to a maximum and then drops off to zero rather than decreasing monotonically. The maximum shifts towards zero value of time for lower value of $C$ with the maximum in anisotropy increasing very sharply. This is an interesting feature arising out of the tachyon field behaviour in the low energy limiting case where the field could not be obtained in general but under a special condition.  It is to be noted that the initial anisotropy starts off with a much lower value in the high energy limit of an universe evolving from a singularity. 

The value is considerably many orders of magnitude lower in this case for the integration constant $B_{-}$.  For a non-singular universe the initial anisotropies have identical initial values in both the energy limits as expected in the absence of an initial singularity. For positive $\eta$ the initial value of anisotropy parameter is exactly identical for GR and brane in the case $C=1$ and for negative $\eta$ also it is almost identical in both the limiting energy cases. However, in the brane limiting case the value of $C$ has to be lowered considerably as compared to the other cases in order to get any difference in the behaviour of vanishing of anisotropy. For universe evolving from singularity initial value of anisotropy is few orders of magnitude lower in the brane limit for positive $\eta$ and many orders of magnitude lower in the brane limit than in GR limit for negative $\eta$. For a singular universe, the difference in initial anisotropies is more apparent in the limiting energy cases as well as difference in sign of parameter $\sqrt{C}$ associated with the tachyon field. 

Thus, one can note that in the present study we are able to show that evolution from anisotropy to isotropy is attained via inflation with a tachyon field for both universes evolving from an initial singularity and non-singular.

\section*{Acknowledgments}  BCP and SR are thankful to the Inter-University Centre for Astronomy and Astrophysics (IUCAA), Pune, India for providing Visiting Associateship under which a part of this work was carried out. RS and PP are grateful to the IRC, North Bengal University for allowing a short term visit with all the working facilities.

\end{document}